\documentclass[runningheads]{llncs}
\usepackage{graphicx}
\usepackage{epstopdf}
\usepackage{cite}
\usepackage{amsfonts,amssymb,amsmath}
\usepackage{epsf}
\usepackage{color}
\usepackage{multirow}
\usepackage{longtable}
\usepackage{booktabs}
\usepackage[table]{xcolor}
\usepackage{ragged2e}
\usepackage{subfigure}
\usepackage{marvosym}
\usepackage{rotating}
\usepackage{footnote}
\usepackage{floatrow}
\usepackage{hyperref}
\hypersetup{
colorlinks=true,
linkcolor=cyan,
filecolor=blue,      
urlcolor=red,	
citecolor=green,
}
\newfloatcommand{capbtabbox}{table}[][\FBwidth]
\begin{document}
\title{Reliable Multimodality Eye Disease Screening via Mixture of Student's t Distributions}
\author{Ke Zou\inst{1,2,*}, Tian Lin\inst{3,4,*}, Xuedong Yuan\inst{1,2}\textsuperscript{\Letter}, Haoyu Chen\inst{3,4}\textsuperscript{\Letter},\\ Xiaojing Shen\inst{1,5}, Meng Wang\inst{6}, Huazhu Fu\inst{6} }
\authorrunning{K. Zou et al.}

\institute{National Key Laboratory of Fundamental Science on Synthetic Vision,\\ Sichuan
University, Sichuan, China \and College of Computer Science, Sichuan University, Sichuan, China \and Joint Shantou International Eye Center, Shantou University and the Chinese University of Hong Kong, Guangdong, China \and
Medical College, Shantou University, Guangdong, China \and College of Mathematics, Sichuan University, Sichuan, China \and Institute of High Performance Computing, A*STAR, Singapore\\
\email{yxd@scu.edu.cn; drchenhaoyu@gmail.com}}
\renewcommand{\thefootnote}{}
\footnotetext{* denotes equal contribution.}

%
\maketitle              
\begin{abstract}
Multimodality eye disease screening is crucial in ophthalmology as it integrates information from diverse sources to complement their respective performances. However, the existing methods are weak in assessing the reliability of each unimodality, and directly fusing an unreliable modality may cause screening errors. To address this issue, we introduce a novel multimodality evidential fusion pipeline for eye disease screening, EyeMoS$t$, which provides a measure of confidence for unimodality and elegantly integrates the multimodality information from a multi-distribution fusion perspective. Specifically, our model estimates both local uncertainty for unimodality and global uncertainty for the fusion modality to produce reliable classification results. More importantly, the proposed mixture of Student's $t$ distributions adaptively integrates different modalities to endow the model with heavy-tailed properties, increasing robustness and reliability. Our experimental findings on both public and in-house datasets show that our model is more reliable than current methods. Additionally, EyeMos$t$ has the potential ability to serve as a data quality discriminator, enabling reliable decision-making for multimodality eye disease screening.
\end{abstract}

\section{Introduction}
Retinal fundus images and Optical Coherence Tomography (OCT) are common 2D and 3D imaging techniques used for eye disease screening. Multimodality learning usually provides more complementary information than unimodality learning~\cite{zhou2019review,cai2022uni4eye, cai2022corolla}. This motivates researchers to integrate multiple modalities to improve the performance of eye disease screening. Current multimodality learning methods can be roughly classified into early, intermediate, and late fusion, depending on the fusion stage~\cite{2018multimodalreview}.
For multimodality ophthalmic image learning, recent works have mainly focused on the early fusion~\cite{hua2020convolutional,li2020self,rodrigues2020element} and intermediate fusion stages~\cite{wang2022learning,cai2022uni4eye, cai2022corolla, li2022multimodal}. Early fusion-based approaches integrate multiple modalities directly at the data level, usually by concatenating the raw or preprocessed multimodality data. Hua \textit{et al.}~\cite{hua2020convolutional} combined preprocessed fundus images and wide-field swept-source optical coherence tomography angiography at the early stage and then extracted representational features for diabetic retinopathy recognition. Intermediate fusion strategies allow multiple modalities to be fused at different intermediate layers of the neural networks. He \textit{et al.}~\cite{he2021multi} extracted different modality features with convolutional block attention module~\cite{woo2018cbam} and modality-specific attention mechanisms, then concatenated them to realize the multimodality fusion for retinal image classification. However, few studies have explored multimodality eye disease screening at the late fusion stage. Furthermore, the above methods do not adequately assess the reliability of each unimodality, and may directly fuse an unreliable modality with others. This could lead to screening errors and be challenging for real-world clinical safety deployment. To achieve this goal, we propose a reliable framework for the multimodality eye disease screening, which provides a confidence (uncertainty) measure for each unimodality and adaptively fuses multimodality predictions in principle.

Uncertainty estimation is an effective way to provide a measure of reliability for ambiguous network predictions. The current uncertainty estimation methods mainly include Bayesian neural networks, deep ensemble methods, and deterministic-based methods. Bayesian neural networks~\cite{mackay1992practical,neal2012bayesian,ranganath2014black} learn the distribution of network weights by treating them as random variables. However, these methods are affected by the challenge of convergence and have a large number of computations. The dropout method has alleviated this issue to a certain extent~\cite{17dropoutCV}. Another uncertainty estimation way is to learn an ensemble of deep networks~\cite{ensemble17}. 
Recently, to alleviate computational complexity and overconfidence~\cite{evidential18}, deterministic-based methods~\cite{evidential18,zou2023evidencecap,malinin2018predictive,ICMLdeterministic20,2020NIPSdeterministic} have been proposed to directly output uncertainty in a single forward pass through the network. For multimodal uncertainty estimation, the Trusted Multi-view Classification (TMC)~\cite{2020trusted} is a representative method that proposes a new paradigm of multi-view learning by dynamically integrating different views at the evidence level. However, TMC has a limited ability to detect Out-Of-Distribution (OOD) samples~\cite{junguncertainty}. This attributes to TMC is particularly weak in modeling epistemic uncertainty for each single view~\cite{17dropoutCV}. Additionally, the fusion rule in TMC fails to account for conflicting views, making it unsuitable for safety-critical deployment~\cite{zadeh1984review}. To address these limitations, we propose EyeMoSt, a novel evidential fusion method that models both aleatoric and epistemic uncertainty in unimodality, while efficiently integrating different modalities from a multi-distribution fusion perspective.

In this work, we propose a novel multimodality eye disease screening method, called EyeMoS$t$, that conducts Fundus and OCT modality fusion in a reliable manner. Our EyeMoS$t$ places Normal-inverse Gamma (NIG) prior distributions over the pre-trained neural networks to directly learn both aleatoric and epistemic uncertainty for unimodality. Moreover, Our EyeMoSt introduces the Mixture of Student's $t$ (MoS$t$) distributions, which provide robust classification results with global uncertainty. More importantly, MoS$t$ endows the model with robustness under heavy-tailed property awareness. We conduct sufficient experiments on two datasets for different eye diseases (\textit{e.g.}, glaucoma grading, age-related macular degeneration, and polypoid choroidal vasculopathy) to verify the reliability and robustness of the proposed method. In summary, the key contributions are as follows:\\
1) We propose a novel multimodality eye disease screening method, EyeMoSt, which conducts reliable fusion of Fundus and OCT modalities.\\
2) Our EyeMoSt introduces the MoSt distributions, which provide robust classification results with local and global uncertainty.\\
3) We conduct extensive experiments on two datasets for different eye diseases.\footnote[1]{Our code has been released in \url{https://github.com/Cocofeat/EyeMoSt}. }

\begin{figure}[!t]
\centering
\includegraphics[width=1\linewidth]{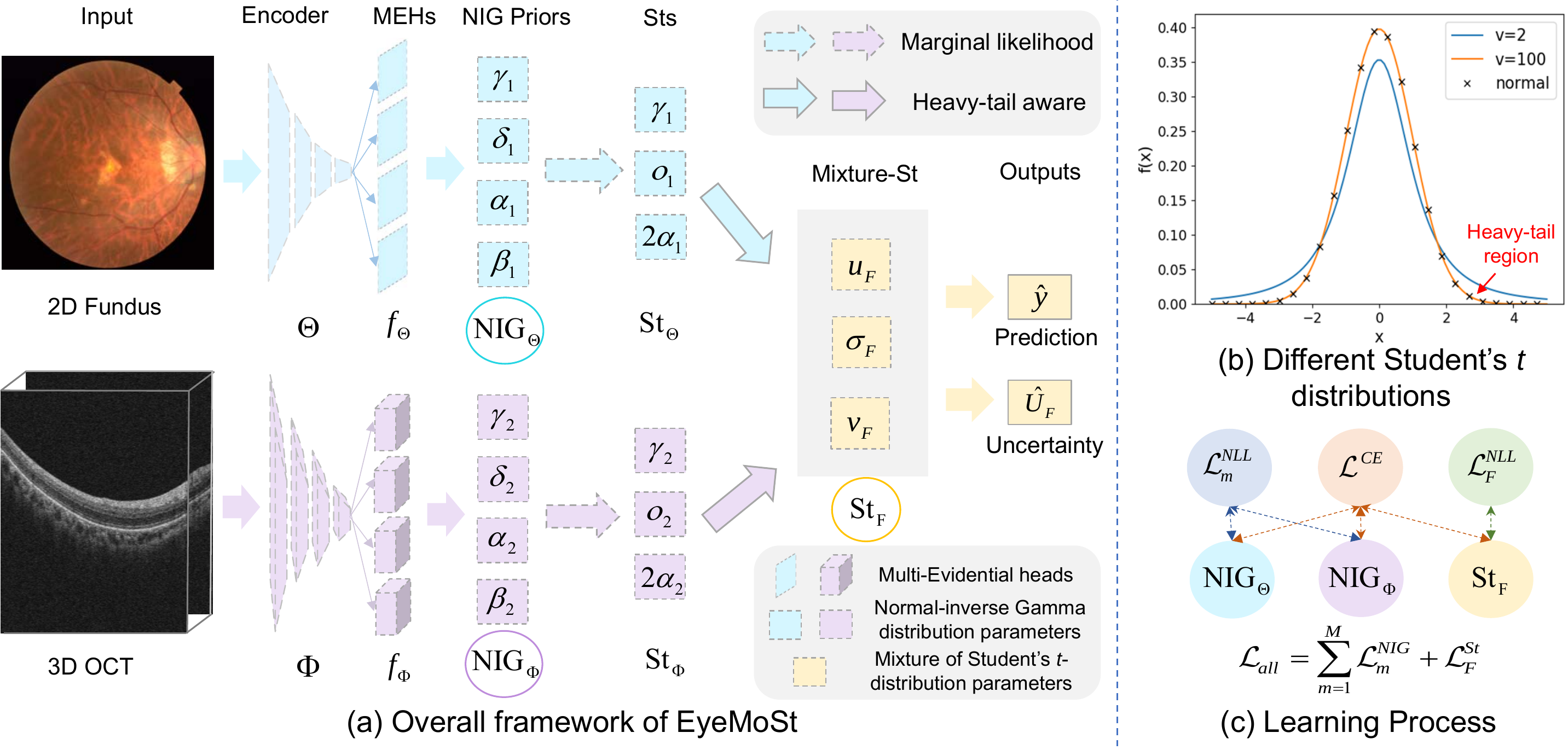}
\caption{Reliable multimodality Eye Disease Screening pipeline. (a) Overall framework of EyeMoS$t$. (b) Student's $t$ Distributions with different degrees of freedom. (c) The overall learning process of EyeMoS$t$. }
\label{F_1}
\end{figure}

\section{Method}
In this section, we introduce the overall framework of our EyeMoS$t$, which efficiently estimates the aleatoric and epistemic uncertainty for unimodality and adaptively integrates Fundus and OCT modalities in principle. As shown in Fig.~\ref{F_1} (a),  we first employ the 2D/3D neural network encoders to capture different modality features. Then, we place multi-evidential heads after the trained networks to model the parameters of higher-order NIG distributions for unimodality. To merge these predicted distributions, We derive the posterior predictive of the NIG distributions as Student's $t$ (S$t$) distributions. Particularly, the Mixture of Student's $t$ (MoS$t$) distributions is introduced to integrate the distributions of different modalities in principle. Finally, we elaborate on the training pipeline for the model evidence acquisition.

Given a multimodality eye dataset $ {\mathcal D} = \left\{ {\left\{ {{\bf{x}}_m^i} \right\}_{m = 1}^M} \right\}$ and the corresponding label ${y^i}$, the intuitive goal is to learn a function that can classify different categories. Fundus and OCT are common imaging modalities for eye disease screening. Therefore, here M=2, ${{\bf{x}}_1^i}$ and ${{\bf{x}}_2^i}$ represent Fundus and OCT input modality data, respectively. We first train 2D encoder ${\rm{\Theta }}$ of Res2Net~\cite{Res2Net}  and 3D encoder ${\rm{\Phi}}$ of MedicalNet~\cite{Med3D2019} to identify the feature-level informativeness, which can be defined as ${\Theta \left( {{\bf{x}}_1^i} \right) }$ and ${{\rm{\Phi}} \left( {{\bf{x}}_2^i} \right) }$, respectively.

\subsection{Uncertainty estimation for unimodality} 
We extend the deep evidential regression model~\cite{2020deepregression} to multimodality evidential classification for eye disease screening. To this end, to model the uncertainty for Fundus or OCT modality, we assume that the observe label $y^i$ is drawn from a Gaussian ${\mathcal N}\left( {{y^i}|\mu,{\sigma ^2}} \right)$, whose mean and variance are governed by an evidential prior named the NIG distribution: 
\begin{equation}
\label{E_1}
{\rm{NIG}}\left( {\mu ,{\sigma ^2}|{{\bf{p}}_m}} \right) = {\mathcal N}\left( {\mu |{\gamma _m},\frac{{{\sigma ^2}}}{{{\delta _m}}}} \right){{\rm\Gamma ^{ - 1}}}\left( {{\sigma ^2}|{\alpha _m},{\beta _m}} \right),
\end{equation}
where $\rm{\Gamma ^{ - 1}}$ is an inverse-gamma distribution, ${\gamma _m} \in \mathbb{R} ,{\delta _m} > 0,{\alpha _m} > 1,{\beta _m} > 0$ are the learning parameters.  Specifically, the multi-evidential heads will be placed after the encoders ${\rm{\Theta }}$ and ${\rm{\Phi}}$ (as shown in Fig.~\ref{F_1} (a)), which outputs the prior NIG parameters $ {{\bf{p}}_m} = \left( {{\gamma _m},{\delta _m},{\alpha _m},{\beta _m}} \right) $. As a result, the aleatoric (AL) and epistemic (EP) uncertainty can be estimated by the $\mathbb{E}\left[ {{\sigma ^2}} \right]$ and the ${\rm{Var}}\left[ \mu  \right]$, respectively, as:
\begin{equation}
\label{E_2}
{\rm{AL}} = {\rm{E}}\left[ {{\sigma ^2}} \right] = \frac{{{\beta _m}}}{{{\alpha _m} - 1}}, \quad \quad {\rm{E}}{{\rm{P}}} = {\rm{Var}}\left[ \mu  \right] = \frac{{{\beta _m}}}{{{\delta _m}\left( {{\alpha _m} - 1} \right)}}.
\end{equation}
Then, given the evidence distribution parameter ${\bf{p}}_m$, the marginal likelihood is calculated by marginalizing the likelihood parameter: 
\begin{equation}
\label{E_3} 
p\left( {{y^i}|x_{_m}^i,{{\bf{p}}_m}} \right) = \int_\mu  {\int_{{\sigma ^2}} {p\left( {{y^i}|x_{_m}^i,\mu ,{\sigma ^2}} \right){\rm{NIG}}\left( {\mu ,{\sigma ^2}|{{\bf{p}}_m}} \right){\rm{d}}} } \mu {\rm{d}}{\sigma ^2}. 
\end{equation}
Interacted by the prior and the Gaussian likelihood of each unimodality~\cite{2020deepregression}, its analytical solution does exist and yields an S$t$ prediction distribution as:
\begin{align}
\label{E_4} 
p\left( {{y^i}|x_{_m}^i,{{\bf{p}}_m}} \right) &= \frac{{{\rm{\Gamma }}\left( {{\alpha _m} + \frac{1}{2}} \right)}}{{{\rm{\Gamma }}\left( {{\alpha _m}} \right)}}\sqrt {\frac{{{\delta _m}}}{{2\pi {\beta _m}\left( {1 + {\delta _m}} \right)}}} {\left( {1 + \frac{{{\delta _m}{{\left( {{y^i} - {\gamma _m}} \right)}^2}}}{{2{\beta _m}\left( {1 + {\delta _m}} \right)}}} \right)^{ - \left( {{\alpha _m} + \frac{1}{2}} \right)}} \nonumber \\
&= St \left( {{y^i};{\gamma _m},{o_m},2{\alpha _m}} \right),  
\end{align}
with ${o_m} = \dfrac{{{\beta _m}\left( {1 + {\delta _m}} \right)}}{{{\delta _m}{\alpha _m}}}$. The complete derivations of Eq.~\ref{E_4} are available in Supplementary S1.1. Thus, the two modalities distributions are transformed into the student's $t$ Distributions $S{\rm{t}}\left( {{y^i};{u_m},{\Sigma_m},{v_m}} \right) = S{\rm{t}}\left( {{y^i};{\gamma _m},{o_m},2{\alpha _m}} \right)$, with ${u_m} \in \mathbb{R},{\Sigma_m} > 0,{v_m} > 2$.

\subsection{Mixture of Student's $t$ Distributions (MoS$t$)}
Then, we focus on fusing multiple S$t$ Distributions from different modalities. How to rationally integrate multiple S$t$s into a unified S$t$ is the key issue. To this end, the joint modality of distribution can be denoted as:
\begin{equation}
\label{E_5}
S{\rm{t}}\left( {{y^{_i}};{u_F},{\Sigma_F},{v_F}} \right) = S{\rm{t}}\left( {{y^{_i}};{{\left[ {\begin{array}{*{20}{c}}
{u_{_1}^i}\\
{u_{_2}^i}
\end{array}} \right]}},\left[ {\begin{array}{*{20}{c}}
{{\Sigma _{11}}}&{{\Sigma _{12}}}\\
{\Sigma _{_{21}}}&{{\Sigma _{22}}}
\end{array}} \right],{{\left[ {\begin{array}{*{20}{c}}
{v_{_1}^i}\\
{v_{_2}^i}
\end{array}} \right]}}} \right).
\end{equation}
In order to preserve the closed S$t$ distribution form and the heavy-tailed properties of the fusion modality, the updated parameters are given by~\cite{roth2013student}. In simple terms, we first adjust the degrees of freedom of the two distributions to be consistent. As shown in Fig.~\ref{F_1} (b), the smaller values of degrees of freedom (DOF) $v$ has heavier tails. Therefore, we construct the decision value ${\tau _m} = {v_m}$ to approximate the parameters of the fused distribution. We assume that multiple S$t$ distributions are still an approximate S$t$ distribution after fusion. Assuming that the degrees of freedom of $\tau _1$ are smaller than $\tau _2$, then, the fused S$t$ distribution $S{\rm{t}}\left( {{y^{_i}};u_{_F},\Sigma _{_F},v_{_F}} \right)$ will be updated as:
\begin{equation}
\label{E_6}
{v_F}{\rm{ = }}{v_1},\quad \quad {u_F}{\rm{ = }}{u_1}, \quad \quad {\Sigma _F}{\rm{ = }}\frac{1}{2}\left( {{\Sigma _1} + \frac{{{v_2}\left( {{v_1} - 2} \right)}}{{{v_1}\left( {{v_2} - 2} \right)}}{\Sigma _2}} \right).
\end{equation}
More intuitively, the above formula determines the modality with a stronger heavy-tailed attribute. That is, according to the perceived heavy-tailed attribute of each modality, the most robust modality is selected as the fusion modality. Finally, the prediction and uncertainty of the fusion modality is given by:
\begin{equation}
\label{E_7}
{{\hat y}^i} = \mathbb{E}{_{p\left( {x_F^i,{{\bf{p}}_F}} \right)}}\left[ {{y^i}} \right]= {u_F},\quad {{\hat U}_F} = {\mathbb{E}}\left[ {\sigma _F^2} \right] = {\Sigma _F}\frac{{{v_F}}}{{{v_F} - 2}}.
\end{equation}

\subsection{Learning the evidential distributions} 
Under the evidential learning framework, we expect more evidence to be collected for each modality, thus, the proposed model is expected to maximize the likelihood function of the model evidence. Equivalently, the model is expected to minimize the negative log-likelihood function, which can be expressed as:
\begin{align} 
\label{E_9} 
{\mathcal L}_m^{NLL} =& \log \frac{{\Gamma \left( {{\alpha _m}} \right)\sqrt {\frac{\pi }{{{\delta_m}}}} }}{{\Gamma \left( {{\alpha _m} + \frac{1}{2}} \right)}} - {\alpha _m}\log \left( {2{\beta _m}\left( {1 + {\delta_m}} \right)} \right) \nonumber \\
&+ \left( {{\alpha _m} + \frac{1}{2}} \right)\log \left( {{{\left( {y^{_i} - {\gamma _m}} \right)}^2}{\delta_m} + 2{\beta _m}\left( {1 + {\delta_m}} \right)} \right).
\end{align}
Then, to fit the classification tasks, we introduce the cross entropy term ${\mathcal L}_m^{CE}$: 
\begin{equation}
\label{E_10}
{\mathcal L}_m^{NIG} = {\mathcal L}_m^{NLL} + {\lambda}{\mathcal L}_m^{CE},
\end{equation}
where ${\lambda}$ is the balance factor set to 0.5. For further information on the selection of hyperparameter $\lambda$, please refer to Supplementary S2. Similarly, for the fusion modality, we first maximize the likelihood function of the model evidence as follows:
\begin{equation}
\label{E_11}
{\mathcal L}_F^{NLL} = {\rm{log}}\Sigma_F  + \log \frac{{\Gamma \left( {\frac{v_F}{2}} \right)}}{{\Gamma \left( {\frac{{v_F + 1}}{2}} \right)}} + \log \sqrt {v_F\pi }  + \frac{\left( {v_F + 1} \right)}{2}\log \left( {1 + \frac{{{{\left( {y^{_i} - u_F} \right)}^2}}}{{v_F{\Sigma_F}}}} \right),
\end{equation}
Complete derivations of Eq.~\ref{E_11} are available in Supplementary S1.2. Then, to achieve better classification performance, the cross entropy term ${\mathcal L}_m^{CE}$ is also introduced into Eq.~\ref{E_11} as below:
\begin{equation}
\label{E_12}
{\mathcal L}_F^{St} = {\mathcal L}_F^{NLL} + {\lambda}{\mathcal L}_F^{CE},
\end{equation}
Totally, the evidential learning process for multimodality screening can be denoted as:
\begin{equation}
\label{E_13}
{\mathcal L}_{all} = \sum\limits_{m = 1}^M{\mathcal L}_m^{NIG} + {\mathcal L}_F^{St}.
\end{equation}
In this paper, we mainly consider the fusion of two modalities, $M=2$.

\section{Experiments}

\noindent\textbf{Datasets:} In this paper, we verify the effectiveness of EyeMoS$t$ on the two datasets. For the glaucoma recognition, We validate the proposed method on the GAMMA~\cite{wu2022gamma} dataset. It contains 100 paired cases with a three-level glaucoma grading. They are divided into the training set and test set with 80 and 20 respectively. We conduct the five-fold cross-validation on it to prevent performance improvement caused by accidental factors. Then, we test our method on the in-house collected dataset, which includes Age-related macular degeneration (AMD) and polypoid choroidal vasculopathy (PCV) diseases. They are divided into training, validation, and test sets with 465, 69, and 70 cases respectively. More details of the dataset can be found in Supplementary S2. Both of these datasets are including the paired cases of Fundus (2D) and OCT (3D). \footnote{The ethical approval of this dataset was obtained from the Ethical Committee.} 

\noindent\textbf{Training Details:} Our proposed method is implemented in PyTorch and trained on NVIDIA GeForce RTX 3090. Adam optimization~\cite{Adam14} is employed to optimize the overall parameters with an initial learning rate of 0.0001. The maximum of epoch is 100. The data augmentation techniques for GAMMA dataset are similar to~\cite {cai2022corolla}, including random grayscaling, random color jitter, and random horizontal flipping. All inputs are uniformly adjusted to $256\times256$ and $128\times256\times128$ for Fundus and OCT modalities. The batch size is 16. 

\noindent\textbf{Compared Methods \& Metrics:} We compare the following six methods: For different fusion stage strategies,  \textbf{a) B-EF} Baseline of the early fusion~\cite{hua2020convolutional} strategy, \textbf{b) B-IF}  Baseline of the intermediate typical fusion method, \textbf{c) $M^2$LC}~\cite{woo2018cbam} of the intermediate fusion method and the later fusion method \textbf{d) TMC}~\cite{2020trusted} are used as comparisons. B-EF is first integrated at the data level, and then passed through the same MedicalNet~\cite{Med3D2019}. B-IF first extracts features by the encoders (same with us), and then concatenates their output features as the final prediction. For the uncertainty quantification methods, \textbf{e) MCDO} (Monte Carlo Dropout) employs the test time dropout as an approximation of a Bayesian neural network~\cite{16dropout}. \textbf{f) DE} (Deep ensemble) quantifies the uncertainties by ensembling multiple models~\cite{ensemble17}. We adopt the accuracy (ACC) and Kappa metrics for intuitive comparison with different methods. Particularly, expected calibration error (ECE)~\cite{maronas2020calibration} is used to compare the calibration of the uncertainty algorithms. 

\begin{table}[!t]
  \centering
  \label{T_1}
  \caption{Comparisons with different algorithms on the GAMMA and  in-house dataset. F and O denote the Fundus and OCT modalities, respectively. The top-2 results are highlighted in \textcolor{red}{Red} and \textcolor{blue}{Blue}. Higher ACC and Kappa, and Lower ECE mean better. }
  \resizebox{1\textwidth}{!}{
    \begin{tabular}{cccccccccccccc}
    \toprule
    \multicolumn{2}{c}{\multirow{4}[6]{*}{Methods}} & \multicolumn{6}{c}{GAMMA dataset}             & \multicolumn{6}{c}{In-house dataset} \\
    \multicolumn{2}{c}{} & \multicolumn{2}{c}{\multirow{2}[3]{*}{Original}} & \multicolumn{4}{c}{Gaussian noise} & \multicolumn{2}{c}{\multirow{2}[3]{*}{Original}} & \multicolumn{4}{c}{Gaussian noise} \\
\cmidrule{5-8}\cmidrule{11-14}    \multicolumn{2}{c}{} & \multicolumn{2}{c}{} & \multicolumn{2}{c}{$\sigma$=0.1 (F)} & \multicolumn{2}{c}{$\sigma$=0.3  (O)} & \multicolumn{2}{c}{} & \multicolumn{2}{c}{$\sigma$=0.1 (F)} & \multicolumn{2}{c}{$\sigma$=0.3 (O)} \\
\cmidrule{5-8}\cmidrule{11-14}    \multicolumn{2}{c}{} & ACC   & Kappa & ACC   & Kappa & ACC   & Kappa & ACC   & ECE   & ACC   & ECE   & ACC   & ECE \\
    \midrule
    \multicolumn{2}{c}{B-IF} & 0.700  & 0.515  & 0.623  & 0.400  & 0.530  & 0.000  & 0.800  & 0.250 & 0.593  &  0.450 & 0.443  & 0.850  \\
    \multicolumn{2}{c}{B-EF\cite{hua2020convolutional}} & 0.660  & 0.456  & 0.660  & 0.452  & 0.500  & 0.000  & 0.800  & 0.200 & 0.777 & 0.223 & 0.443  & 0.557  \\
    \multicolumn{2}{c}{$M^2$LC\cite{woo2018cbam}} & 0.710  & 0.527  & 0.660  & 0.510  & 0.500  & 0.000  & 0.814  & 0.186 & 0.786 & 0.214 & 0.443  & 0.557  \\
    \multicolumn{2}{c}{MCDO\cite{16dropout}} & 0.758  & 0.636  & 0.601  & 0.341  & 0.530  & 0.000  & 0.786  & 0.214  & 0.771  & 0.304  & 0.429  & 0.571  \\
    \multicolumn{2}{c}{DE\cite{ensemble17}} & 0.710  & 0.539 & 0.666  & 0.441 & 0.530  & 0.000  & 0.800   & 0.200   & 0.800   & 0.200   & 0.609 & 0.391 \\
    \multicolumn{2}{c}{TMC\cite{2020trusted}} & 0.810  & 0.658  & 0.430  & 0.124  & 0.550  & 0.045  & \textcolor{red}{0.829}  & \textcolor{red}{0.171}  & 0.814  & 0.186  & 0.443  & 0.557  \\
    \multicolumn{2}{c}{Our} & \textcolor{red}{0.850}  & \textcolor{red}{0.754}  & \textcolor{blue}{0.663} & \textcolor{blue}{0.458}& \textcolor{red}{0.830} & \textcolor{red}{0.716}  & \textcolor{red}{0.829}& \textcolor{red}{0.171} & \textcolor{blue}{0.800} & \textcolor{blue}{0.200}& \textcolor{red}{0.829}  & \textcolor{red}{0.171} \\
    \bottomrule
    \end{tabular}}%
  \label{tab:addlabel}%
\end{table}%

\begin{figure}[!t]
\centering
\includegraphics[width=1\linewidth]{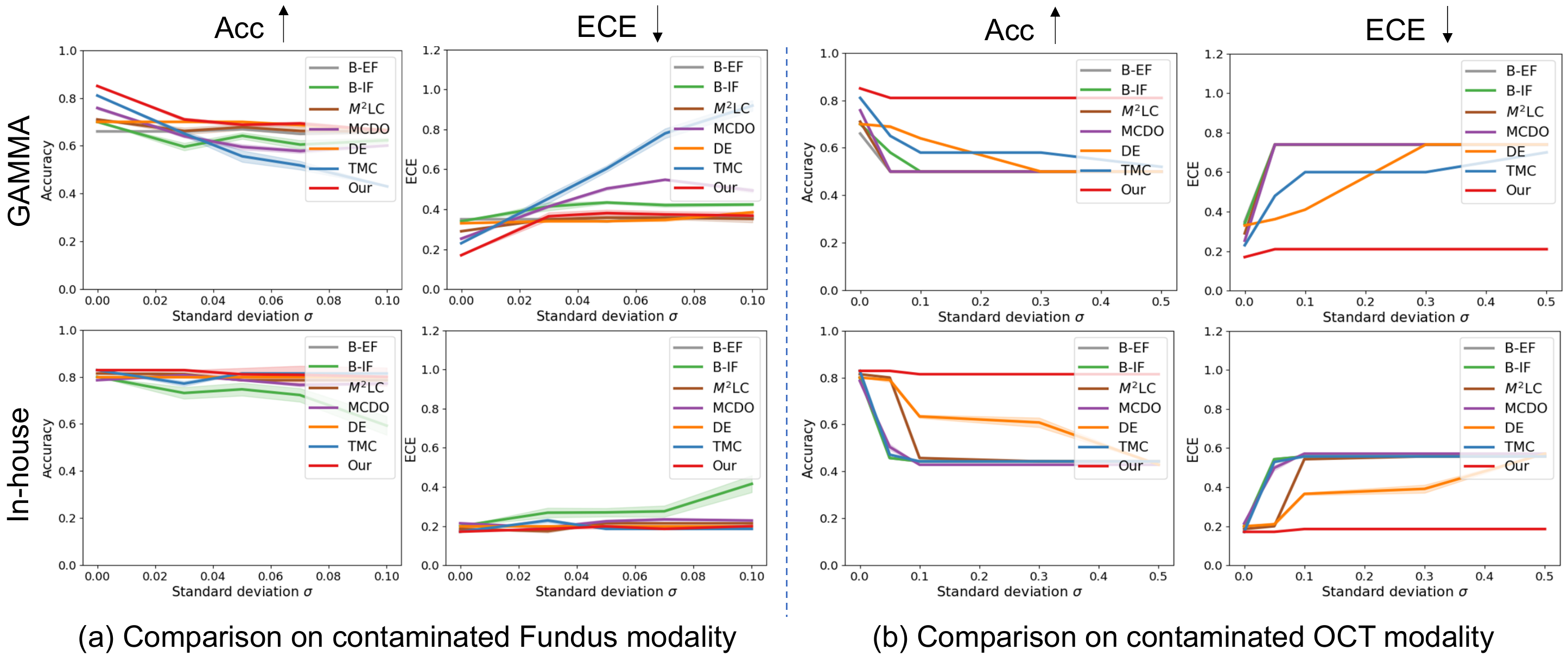}
\caption{Accuracy and ECE performance of different algorithms in contaminated single modality with different levels of noise on GAMMA and in-house datasets.}
\label{F_3}
\end{figure}

\noindent\textbf{Comparison and Analysis:} 
We reported our algorithm with different methods on the GAMMA and in-house datasets in Tab.~\ref{T_1}. First, we compare these methods under the clean multimodality eye data. Our method obtained competitive results in terms of ACC and Kappa. Then, to verify the robustness of our model, we added Gaussian noise to Fundus or OCT modality ($\sigma=0.1/0.3$) on the two datasets. Compared with other methods, our EyeMoS$t$ maintains classification accuracy in noisy OCT modality, while comparable in noisy Fundus modality. More generally, we added different Gaussian noises to Fundus or OCT modality, as shown in Fig.~\ref{F_3}. The same conclusion can be drawn from Fig.~\ref{F_3}. This is attributed to the perceived long tail in the data when fused. The visual comparisons of different noises to the Fundus/OCT modality on the in-house dataset can be found in Supplementary S2. To further quantify the reliability of uncertainty estimation, we compared different algorithms using the ECE indicator. As shown in Tab.~\ref{T_1} and Fig.~\ref{F_3}, our proposed algorithm performs better in both clean and single pollution modalities. The inference times of the uncertainty-based methods on two modalities on the in-house dataset are 5.01s (MCDO), 8.28s (DE), 3.98s (TMC), and 3.22s (Ours). It can be concluded that the running time of EyeMos$t$ is lower than other methods. In brief, we conclude that our proposed model is more robust and reliable than the above methods.

\begin{figure}[!t]
\centering
\includegraphics[width=1\linewidth]{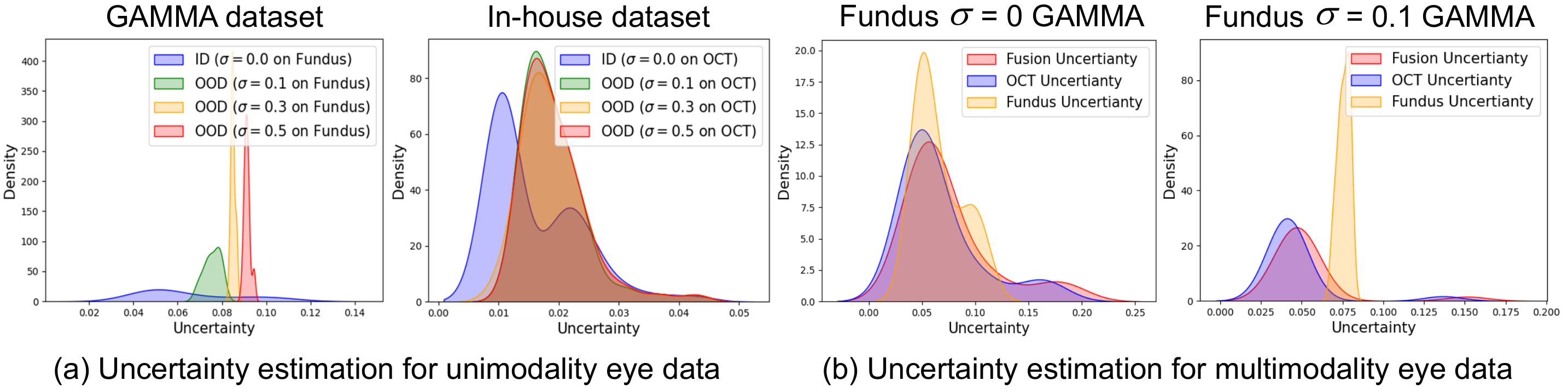}
\caption{Uncertainty density of unimodality and multimodality eye data.}
\label{F_4}
\end{figure}

\noindent\textbf{Understanding uncertainty for unimodality/multimodality eye data:} To make progress towards the multimodality ophthalmic clinical application of uncertainty estimation, we conducted unimodality and multimodality uncertainty analysis for eye data. First, we add more Gaussian noise with varying variances to the unimodality (Fundus or OCT) in the GAMMA and in-house datasets to simulate OOD data. The original samples without noise are denoted as in-distribution (ID) data. Fig.~\ref{F_4} (a) shows a strong relationship between uncertainty and OOD data. Uncertainty in unimodality images increases positively with noise. Here, uncertainty acts as a tool to measure the reliable unimodality eye data. Second, we analyze the uncertainty density of unimodality and fusion modality before and after adding Gaussian noise. As shown in Fig.~\ref{F_4} (b), take adding noise with $\sigma=0.1$ to the Fundus modality on the GAMMA dataset as an example. Before the noise is added, the uncertainty distributions of unimodality and fusion modality are relatively concentrated. After adding noise, the uncertainty distribution of the fusion modality is closer to that of the modality without noise. Hence, EyeMoS$t$ can serve as a tool for measuring the reliable modality in ophthalmic multimodality data fusion. To this end, our algorithm can be used as an out-of-distribution detector and data quality discriminator to inform reliable and robust decisions for multimodality eye disease screening.

\section{Conclusion}
In this paper, we propose the EyeMoS$t$ for reliable and robust screening of eye diseases using evidential multimodality fusion. Our EyeMoS$t$ produces the uncertainty for unimodality and then adaptively fuses different modalities in a distribution perspective. The different NIG evidence priors are employed to model the distribution of encoder observations, which supports the backbones to directly learn aleatoric and epistemic uncertainty. We then derive an analytical solution to the Student’s $t$ distributions of the NIG evidence priors on the Gaussian likelihood function. Furthermore, we propose the MoS$t$ distributions in principle adaptively integrates different modalities, which endows the model with heavy-tailed properties  and is more robust and reliable for eye disease screening. Extensive experiments show that the robustness and reliability of our method in classification and uncertainty estimation on GAMMA and in-house datasets are competitive with previous methods. Overall, our approach has the potential to multimodality eye data discrimitor for trustworthy medical AI decision-making.  In future work, our focus will be on incorporating uncertainty into the training process and exploring the application of reliable multimodality screening for eye diseases in a clinical setting.

\small{\textbf{Acknowledgements:} This work was supported by the National Research Foundation, Singapore under its AI Singapore Programme (AISG Award No: AISG2-TC-2021-003), A*STAR AME Programmatic Funding Scheme Under Project A20H4b0141, A*STAR Central Research Fund, the Science and Technology Department of Sichuan Province (Grant No. 2022YFS0071 \& 2023YFG0273), and the China Scholarship Council (No. 202206240082).}

\bibliography{uncertainty}
\bibliographystyle{splncs04}

\newpage
\begin{appendix}
\section*{Supplementary Materials}
\setcounter{figure}{0}
\setcounter{table}{0}
\setcounter{equation}{0}
\setcounter{page}{1}
\renewcommand{\thetable}{S\arabic{table}}
\renewcommand{\thefigure}{S\arabic{figure}}
\renewcommand{\theequation}{S\arabic{equation}}


\section*{1. Derivations }
\subsection*{1.1. Decomposition details of Eq.~\ref{E_4} }
In this subsection, a Student-t predictive distribution is generated by deriving the posterior predictor from a NIG distribution:\\
\begin{equation}
\begin{array}{l}
p\left( {{y^i}\mid x_m^i,{{\bf{p}}_m}} \right) = \int_\mu  {\int_{{\sigma ^2}} p } \left( {{y^i}\mid x_m^i,\mu ,{\sigma ^2}} \right){\rm{NIG}}\left( {\mu ,{\sigma ^2}\mid {{\bf{p}}_m}} \right){\rm{d}}\mu {\rm{d}}{\sigma ^2}\\
{\rm{ = }}\int_\mu  {\int_{{\sigma ^2}} {\left[ {\sqrt {\frac{1}{{2\pi {\sigma ^2}}}} {e^{ - \frac{{{{\left( {{y^i} - \mu } \right)}^2}}}{{2{\sigma ^2}}}}}} \right]} } \left[ {\frac{{{\beta _m}^\alpha \sqrt {{\delta _m}} }}{{\Gamma \left( {{\alpha _m}} \right)\sqrt {2\pi {\sigma ^2}} }}{{\left( {\frac{1}{{{\sigma ^2}}}} \right)}^{{\alpha _m} + 1}}{e^{ - \frac{{2{\beta _m} + {\delta _m}{{\left( {{\gamma _m} - \mu } \right)}^2}}}{{2{\sigma ^2}}}}}} \right]{\rm{d}}\mu {\rm{d}}{\sigma ^2}\\
 = \int_{{\sigma ^2}} {\frac{{{\beta _m}^\alpha {\sigma ^{ - 2{\alpha _m} - 3}}}}{{\sqrt {2\pi \left( {1 + 1/{\delta _m}} \right)} \Gamma \left( {{\alpha _m}} \right)}}{e^{ - \frac{{2{\beta _m} + \frac{{{\delta _m}{{\left( {{y^i} - {\gamma _m}} \right)}^2}}}{{{\delta _m} + 1}}}}{{2{\sigma ^2}}}}}} {\rm{d}}{\sigma ^2}\\
 = \frac{{\Gamma \left( {{\alpha _m} + \frac{1}{2}} \right)}}{{\Gamma \left( {{\alpha _m}} \right)}}\sqrt {\frac{{{\delta _m}}}{{2\pi {\beta _m}\left( {1 + {\delta _m}} \right)}}} {\left( {1 + \frac{{{\delta _m}{{\left( {{y^i} - {\gamma _m}} \right)}^2}}}{{2{\beta _m}\left( {1 + {\delta _m}} \right)}}} \right)^{ - \left( {{\alpha _m} + \frac{1}{2}} \right)}}\\
 = {\mathop{\rm St}\nolimits} \left( {{y^i};{\gamma _m},{o_m},2{\alpha _m}} \right)
\end{array}
\end{equation}
\subsection*{1.2. Decomposition details of Eq.~\ref{E_11} }\label{APP_1}
For the fusion modality, we expect more evidence to be collected, which is maximizing the likelihood function of the Student's $t$. Equivalently, we minimize the negative log-likelihood function, which can be expressed as: 
\begin{equation}
\label{APP_E1}
p\left( {{y^i};u_F,\Sigma_F ;v_F} \right) =  - \log \left( {\frac{{\Gamma \left( {\frac{{v_F + 1}}{2}} \right)}}{{\Sigma \Gamma \left( {\frac{v_F}{2}} \right)\sqrt {v_F\pi } }}{{\left( {1 + \frac{{{{\left( {{y^i} - u_F} \right)}^2}}}{{v_F{\Sigma_F}}}} \right)}^{ - \frac{1}{2}\left( {v_F + 1} \right)}}} \right),
\end{equation}
Therefore, we train the model to minimize the above equation by:
\begin{equation}
\label{APP_E2}
\begin{array}{l}
{\mathcal L}_F^{NLL}{\rm{ =  log}}\Sigma_F  + \log \frac{{\Gamma \left( {\frac{v_F}{2}} \right)}}{{\Gamma \left( {\frac{{v_F + 1}}{2}} \right)}} + \log \sqrt {v_F\pi }  - \log {\left( {1 + \frac{{{{\left( {{y^i} - u_F} \right)}^2}}}{{v_F{\Sigma_F}}}} \right)^{ - \frac{1}{2}\left( {v_F + 1} \right)}}\\
\quad \quad {\rm{  =  log}}\Sigma_F  + \log \frac{{\Gamma \left( {\frac{v_F}{2}} \right)}}{{\Gamma \left( {\frac{{v_F + 1}}{2}} \right)}} + \log \sqrt {v_F\pi }  + \frac{1}{2}\left( {v_F + 1} \right)\log \left( {1 + \frac{{{{\left( {{y^i} - u_F} \right)}^2}}}{{v_F{\Sigma_F}}}} \right)
\end{array}.
\end{equation}

\section*{2. Experimental details }\label{APP_2}
\begin{figure}
\centering
\includegraphics[width=1\textwidth,height=0.4\textwidth]{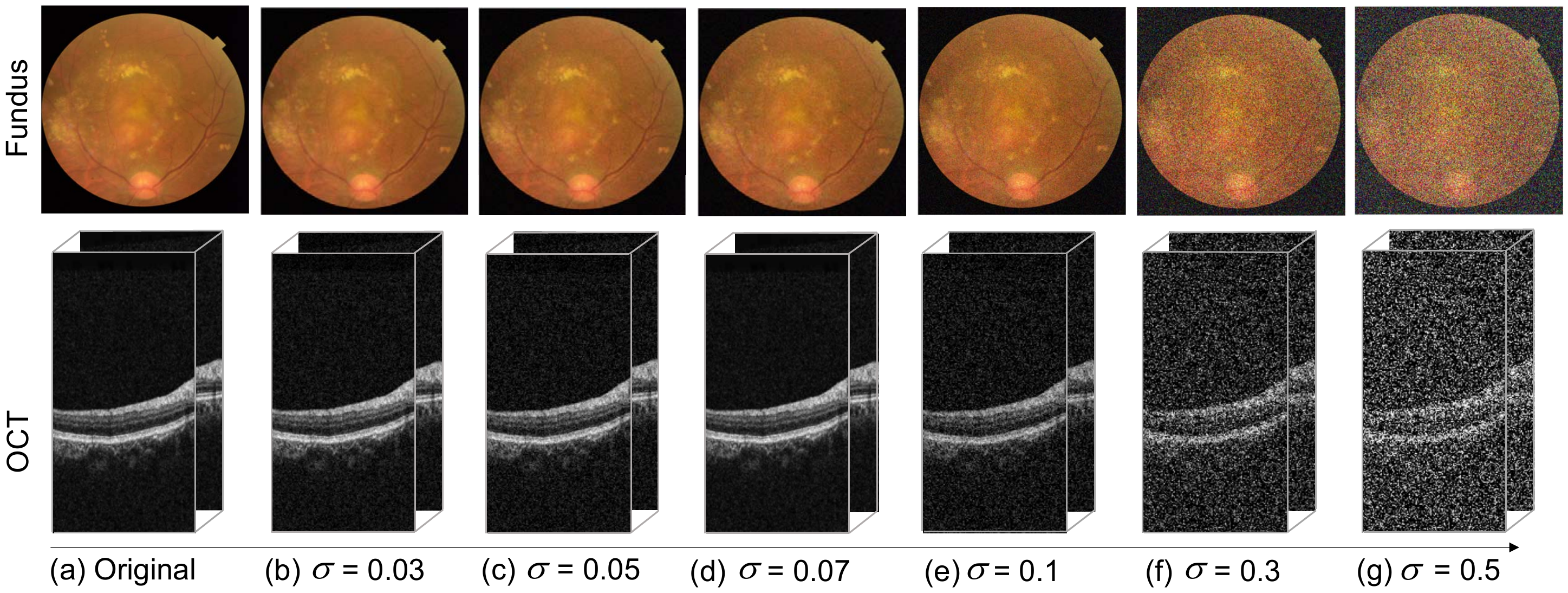}
\caption{Demonstration of different noise to the Fundus/OCT images on the in-house dataset.}
\label{APPF2}
\end{figure}

\textbf{Ablation study.} 1) $\lambda$ is the balance factor between the ${\mathcal L}_m^{NLL}$ loss and the ${\mathcal L}_m^{CE}$ loss. In the experiments below, we demonstrate the importance of augmenting training objective with the evidence classifier loss ${\mathcal L}_m^{CE}$ introduced in EyeMoS$t$. $\lambda  \in \left[ {0,1} \right]$ represents the importance of ${\mathcal L}_m^{CE}$ loss. We performed parameter validation on the in-house dataset. As shown in the Tab.~\ref{APPT_1}, the performance is improved after introducing ${\mathcal L}_m^{CE}$ loss, and the best value is 0.5. 2) Ablation study for overall learning process. Further, we conduct ablation experiments on Eq.~\ref{E_13}, as depicted in Tab.~\ref{APPT_2}. Where B is the baseline of the intermediate typical fusion method. B-IF first extracts features by the encoders (same with us), and then concatenates their output features as the final prediction. ${\mathcal L}_m^{NIG}$ represents pairwise fusion directly after establishing multi-NIG distributions. \\ 
\begin{table}[htbp]
  \centering
  \caption{Parameter selection of $\lambda$ on the in-house dataset}
    \begin{tabular}{ccccccc}
    \toprule
    \label{APPT_1}
    $\lambda$ =  & 0 & 0.1 & 0.2 & 0.5 & 0.7& 1.0 \\
    \midrule
    Acc & 0.800  & 0.771  & 0.814  & \textbf{0.829} & 0.814  & 0.786  \\
    AUC & 0.843  & 0.754  & 0.797  & \textbf{0.850} & 0.850  & 0.810  \\
    F1 &  0.794  & 0.762  & 0.808  & \textbf{0.822} & 0.808  & 0.781  \\
    Recall &  0.800  & 0.771  & 0.814  & \textbf{0.829} & 0.814  & 0.786  \\
    \bottomrule
    \end{tabular}%
\end{table}%
\textbf{Details of in-house dataset.} In-house dataset consists of the 265 AMD samples and 341 PCV samples. The training set, validation set, and test set include pairs of Fundus and OCT of 465, 69, and 70, respectively. The Table~\ref{APPT_3} summarizes the detailed information. 
 The original image size of each Fundus is $2100\times2000$. We adjust it to $256\times256$. The original image size of each OCT is $512\times885$. Each eye case has 128 slices, so the 3D OCT size is adjusted to $128\times256\times128$. As shown in Fig.~\ref{APPF2}, we show some cases with different noise conditions.\\


\begin{table*}
\begin{floatrow}
\capbtabbox{
\resizebox{.95\linewidth}{!}{
    \begin{tabular}{lllrrrr}
    \toprule
    B  & ${\mathcal L}_m^{NIG}$ & ${\mathcal L}_F^{St}$ & \multicolumn{1}{l}{Acc} & \multicolumn{1}{l}{AUC} & \multicolumn{1}{l}{F1} & \multicolumn{1}{l}{Recall} \\
    \midrule
    $\checkmark$ &       &       & 0.800  & 0.854  & 0.792  & 0.800  \\
    $\checkmark$ & $\checkmark$  &   & 0.814  & 0.841 & 0.808  & 0.814 \\
    $\checkmark$ & $\checkmark$  & $\checkmark$   & 0.829  & 0.850  & 0.822  & 0.829  \\
    \bottomrule
    \end{tabular}%
    }
}{
 \caption{The result of ablation Study.}
 \label{APPT_2}
}
\capbtabbox{
\resizebox{.95\linewidth}{!}{
    \begin{tabular}{ccccc}
    \toprule
    In-house & Train & Validation  & Test & Total \\
    \midrule
    AMD & 205   & 27   & 31  & 265\\
    PCV & 260    & 42    & 39 & 341 \\
    All  & 465    & 69    & 70 & 604 \\
    \bottomrule
    \end{tabular}%
    }
}{
 \caption{Information of in-house dataset.}
 \label{APPT_3}
 \small
}
\end{floatrow}
\end{table*}

\end{appendix}
\end{document}